\documentclass[aps,prl,showpacs,twocolumn,floatfix,superscriptaddress]{revtex4}
\usepackage{graphicx}
\usepackage{amsmath}
\usepackage{amsfonts}
\usepackage{subfigure}
\begin{document}

\title{Generalized theory for node disruption in finite size complex networks}

\author{Bivas Mitra}
\affiliation{Department of Computer Science and Engineering, Indian
Institute of Technology, Kharagpur 721302, India}
\author{Niloy Ganguly}
\affiliation{Department of Computer Science and Engineering, Indian
Institute of Technology, Kharagpur 721302, India}
\author{Sujoy Ghose}
\affiliation{Department of Computer Science and Engineering, Indian
Institute of Technology, Kharagpur 721302, India}
\author{Fernando Peruani\footnote{Corresponding author: fernando.peruani@cea.fr}} 
\affiliation{CEA-Service de Physique de l'Etat Condens\'{e}, Centre d'Etudes de Saclay, 91191 Gif-sur-Yvette, France}
\affiliation{Institut des Syst\'emes Complexes de Paris \^{I}le-de-France, 57/59, rue Lhomond F-75005 Paris, France}

\date{\today}
\begin{abstract}
After a failure or attack the structure of a complex network changes due to node removal.
Here, we show that the degree distribution of the distorted network, under any node disturbances,
can be easily computed through a simple formula.
Based on this expression, we derive a general condition for the stability of non-correlated finite complex networks under any arbitrary attack.
We apply this formalism to derive an expression for the percolation threshold $f_c$ under a general attack of the form
$f_k \sim k^{\gamma}$, where $f_k$ stands for the probability of a node of degree $k$ of being removed
during the attack.
We show that $f_c$  of a finite network of size $N$ exhibits an additive correction
which scales as $N^{-1}$ with respect to the classical result for infinite networks.
\end{abstract}

\maketitle

The stability of graphs
against various disrupting events is a central issue in the study of complex networks~\cite{callaway2000,new_gen,alb,guill2005,val,tan,liu2005,motter2004,cohen00,cohen01}.
If information is transported across a network, as is the case of
epidemics across social networks or information broadcast through
Internet, the ``damage" of some nodes can dramatically affect the
dynamics of the system.
In the context of disease spreading, this could lead to an epidemic extinction, while in communications to a halt of information broadcast~\cite{vespignani02,colizza07,sneppen05,cohen2003,xia08}.
Both, the topological structure of the network and the nature of the attack determine the resulting effect~\cite{holme02}.
For example, it has been shown that scale-free (SF) networks display a high degree of tolerance against random failures~\cite{cohen00}, while, on the other hand, they are quite
sensitive against intentional attacks~\cite{cohen01}.
Clearly, there are various strategies to perform an intentional attack, and each one of them requires a
different level of knowledge on the network topology~\cite{holme02,wu07,paul07}.
A rather general attack, proposed in~\cite{gallos2004,gallos2005} and that we will use in this paper, takes the form $f_k \sim k^{\gamma}$, where $f_k$ denotes the probability of a node of degree $k$ of being removed during the attack, while $\gamma$ is associated to the degree of knowledge of the attacker.
The analysis of this attack has revealed that in SF networks an increase of $\gamma$ leads to a decrease of the critical fraction of nodes that must be removed to disintegrate the network, i.e., a decrease in the percolation threshold $f_c$~\cite{gallos2004,gallos2005}.
%

Though many results have been derived for infinite SF networks, very little is known about the stability of finite networks.
Typical examples of small size finite networks are ad-hoc networks
of commercial mobile devices, frequently used for
communication~\cite{adhoc1},
temporary peer-to-peer networks formed by BitTorrent clients for
efficient download of file~\cite{BitTorrent} and networks of
autonomous mobile robots~\cite{robots}.
The operation of these systems relies on the robustness of the
highly dynamical underlying network.
Thus, a good understanding on the stability of these small size networks is imperative for these applications.
Moreover, we can say that in general a comprehensive theory for the stability of arbitrary finite networks under any node disturbance is still lacking.
\begin{figure}[b]
\centering\resizebox{5cm}{!}{\rotatebox{0}{\includegraphics{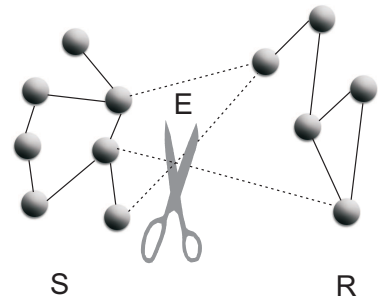}}}
\caption{The scheme illustrates an attack as consisting of two
steps: (a) selection of nodes to be removed (see set R),
and (b) cutting of the edges that run from the surviving nodes (represented by set S) to the set of removed nodes. As the scheme
shows, the attack affects also the degree of the surviving nodes.}
\label{fig:scheme}
\end{figure}

In this paper we attempt to shed some light on this matter by proposing an alternative derivation for the percolation threshold.
Let us clarify that in this paper we are interested in understanding
only the emergence of a giant component at percolation threshold.
Hence, the term {\em percolation} and {\em emergence of giant
component} can be considered synonymous.
In our approach, instead of applying a generating function formalism to find an analytic expression for the percolation threshold as in~\cite{new_gen,callaway2000,wu07}, we used the fact that during an attack the degree distribution of the network changes (see Fig.~\ref{fig:scheme}).
We show that the degree distribution of the distorted (uncorrelated) network, under any node disturbances,
can be easily computed through a simple formula.
Based on this expression, we derive a general condition for the stability of non-correlated complex networks under any arbitrary attack.
This condition applied to the study of network stability under the general attack proposed in~\cite{gallos2004,gallos2005} leads us to a general expression for the percolation threshold $f_c$.
We show that  $f_c$  of a finite network of size $N$ exhibits an additive correction
which scales as $N^{-1}$ with respect to the classical result for infinite networks~\cite{callaway2000, cohen00, cohen01}.
Simulation results confirm all these findings.

\vspace{0.1cm} \noindent {\it Network topology after disturbance.}
A failure or attack can be thought in the following
way. Let $p_k$ be the degree distribution of the network before the
attack.
The first step in the attack is to select the nodes that are going to be removed.
Let us assume that this is performed by means of $f_k$, where $f_k$
represents the probability for a node of degree $k$ being removed
from the network.
Note that the only restriction on $f_k$ is $0 \leq f_k  \leq 1$.
After the node selection, we divide the network into two subsets,
one subset contains the surviving nodes (S) while the other subset
comprises the nodes that are going to be removed (R).
Fig. \ref{fig:scheme} illustrates this procedure. 
At the moment the nodes in R are actually removed, the
degree distribution of the S-nodes is changed due to the removal of the $E$ edges that run between these two subsets.
The probability $\phi$ of finding an edge in 
 subset S that is connected to a node in subset
R is expressed as:
\begin{equation} \label{eq:phi}
\phi =
\frac{\sum_{i=0}^{\infty} {i\,
\,p_i\,f_i}}{\left(\sum_{k=0}^{\infty}{k\,p_k}\right)-1/N }\,.
\end{equation}
The reasoning behind this expression is as follows. The total number of
half-edges in the surviving subset, including the $E$ links that are going
to be removed,  is $\sum_{j=0}^{\infty} {j\,(N\,p_j)\,(1-f_j)}$.
The probability for a randomly chosen half-edge of being removed is simply
$\sum_{i=0}^{\infty} {i\,(N\,p_i)\, f_i}/
(\sum_{k=0}^{\infty} {k\,(N\,p_k)-1})$.
$E$ is the number of half-edges in S times this probability, and
$\phi$ is obtained by dividing $E$ by the number of half-edges in
the subset S.
Notice that the removal of nodes can only lead to a decrease of the degree of a
node.
%
%
Finally, to calculate the degree distribution $p_k'$ after the attack, we
still need to estimate the probability  $p^{s}_{q}$ of finding a
nodes with degree $q$ in the surviving subset S (before cutting the
E edges).
This fraction takes the simple form:
\begin{equation} \label{eq:psk}
p^{s}_{q} = \frac{(1-f_q)p_q}{1-\sum_{i=0}^{\infty}p_i f_i} \,.
\end{equation}
Now we are in condition to compute $p_k'$. Using Eqs. (\ref{eq:phi}) and (\ref{eq:psk}), we obtain the following expression for $p_k'$:
\begin{equation} \label{eq:pk_afterattack}
p_k' = \sum_{q=k}^{\infty} \left( \begin{array}{c} q \\ k \end{array}\right) \phi^{q-k} (1-\phi)^{k}\, p^{s}_{q}\, .
\end{equation}
Eq.~(\ref{eq:pk_afterattack}) can be iteratively evaluated by
replacing $p_k$ with $p_k'$ into Eqs. (\ref{eq:phi}) to
(\ref{eq:pk_afterattack}).
It is instructive to notice that for failure, i.e. $f_k = f$, and assuming $N \gg 1$, Eq. (\ref{eq:phi}) reduces to $\phi = f$
while Eq. (\ref{eq:psk}) becomes $p^{s}_q = p_q$.
In consequence, from Eq. (\ref{eq:pk_afterattack}) we retrieve the degree
distribution $p_k'$ after failure which reads \cite{cohen00}: $ p_k'=\sum_{q=k}^{\infty} \left(
\begin{array}{c} q \\ k \end{array}\right) f^{q-k}
(1-f)^{k} p_q$.
A similar expression has also been used to described $p_k'$ after an ad-hoc
 attack in scale-free networks with $N \gg 1$~\cite{cohen01}.

\begin{figure}
\begin{center}
\includegraphics[angle=0,scale=.65]{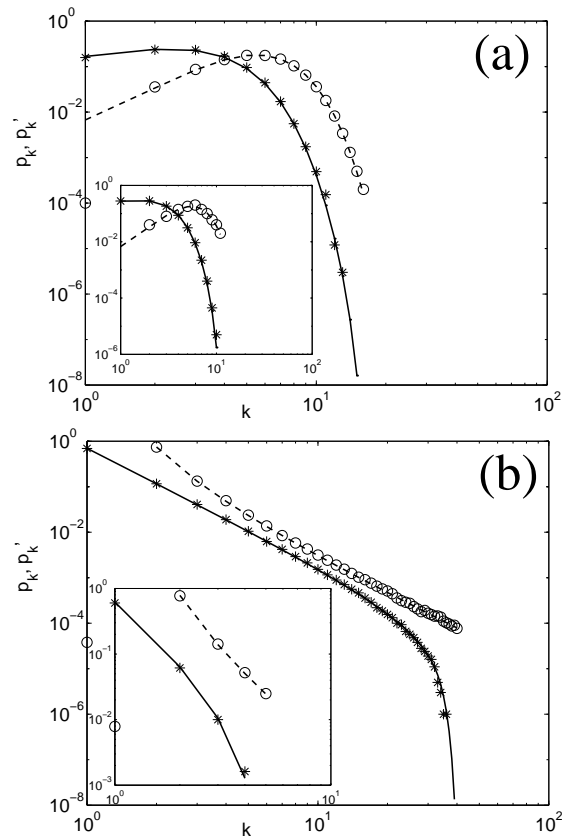}
\end{center}
\caption{Degree distribution before
(circles), and after (stars) the attack. 
In (a) the network topology corresponds to Erd\H{o}s-R\'{e}nyi graphs, with $\langle k
\rangle=5$, while in (b) to scale-free networks, $p_k\sim k^{-\alpha}$ with
$\alpha=2.5$~\cite{powerlaw}.
%
%
Symbols correspond to simulations, solid lines to the theoretical $p_k'$, given by Eq.(\ref{eq:pk_afterattack}), and 
dashed curves to the (theoretical) initial $p_k$.
In (a) and (b) the main figure corresponds to networks with $N=10^5$ nodes, while the inset shows the result for small networks with just $N=50$ nodes. 
} \label{fig:s}
\end{figure}

Fig. \ref{fig:s} shows a comparison between stochastic simulations (symbols) and Eq.~(\ref{eq:pk_afterattack}) (solid line) for two different network topologies, namely
Erd\H{o}s-R\'{e}nyi (ER) graph (a) and scale-free (SF) networks (b)~\cite{powerlaw}.
Removal of nodes (and edges) was performed through an attack of the form $f_k \sim k^{\gamma}$, with $\gamma=1$.
In the figure two different system sizes are shown: $N=10^5$ and $N=50$ (figure insets).
%

\vspace{0.1cm} \noindent {\it Critical condition.}
The following expression tells us whether an infinite network
percolates after an attack~\cite{cohen00}:
\begin{equation} \label{eq:def_percolation}
\kappa' = \frac{\langle k^2 \rangle'}{\langle k \rangle'} > 2 \, ,
\end{equation}
where $\langle k \rangle'$ and $\langle k^2 \rangle'$ refer to the
first and second moments of the degree distribution after the
attack.
We borrow the critical condition for infinite networks given by Eq. (\ref{eq:def_percolation})
to define a ``percolation'' criterion for finite networks. Thus, by definition we assume that the condition $\kappa' = 2$
 determines the point at which the network breaks down~\cite{gallos2004,gallos2005}.
To compute $\langle k \rangle'$ and $\langle k^2 \rangle'$, we utilize the generating function:
\begin{equation} \label{eq:G_0_sum_0}
G_{0}(x) = \sum_{k=0}^{\infty}\sum_{q=k}^{\infty}  \left( \begin{array}{c} q \\ k \end{array}\right) \phi^{q-k} (1-\phi)^{k}   p^s_q \,x^{k} \, .
\end{equation}
After exchanging the order of the sum, the Binomial theorem can be applied, and we obtain:
\begin{equation} \label{eq:G_0_sum}
G_{0}(x) = \sum_{k=0}^{\infty} p^s_q \, \left( (x-1) (1-\phi) + 1\right)^{q} .
\end{equation}
From Eq. (\ref{eq:G_0_sum}), the first two moments can be easily computed as $\langle k \rangle'=dG_{0}(1)/dx$ and $\langle k^2 \rangle'=d^2G_{0}(1)/dx^2+dG_{0}(1)/dx$. After some algebra we obtain that the
critical condition given by Eq. (\ref{eq:def_percolation}) takes the form:
\begin{eqnarray} \nonumber
\left( \sum_k  p_k  (1-f_k) k \right) && \left(\sum_k p_k (1-f_k) k^2 + \sum_k  p_k (f_k -2 ) k \right)  \\
\label{eq:stabilitycond} &&+ \frac{1}{N} \left( \sum_k p_k (1-f_k)(2-k)k \right) = 0 \, .
\end{eqnarray}
Eq.~(\ref{eq:stabilitycond}) determines the stability condition (according to the given definition) for any
uncorrelated network of finite size under any arbitrary attack.
In the limit of $N \to \infty$, Eq. (\ref{eq:stabilitycond}) reduces to:
\begin{equation}\label{eq:infy_N}
\sum_{k=0}^{\infty}p_k k(k(1-f_k)+f_k-2)=0 \,.
\end{equation}
Interestingly, Eq. (\ref{eq:infy_N}) can be also derived through a more classical generating function formalism~\cite{bivas}.
\begin{figure}
\centering\resizebox{\columnwidth}{!}{\rotatebox{0}{\includegraphics{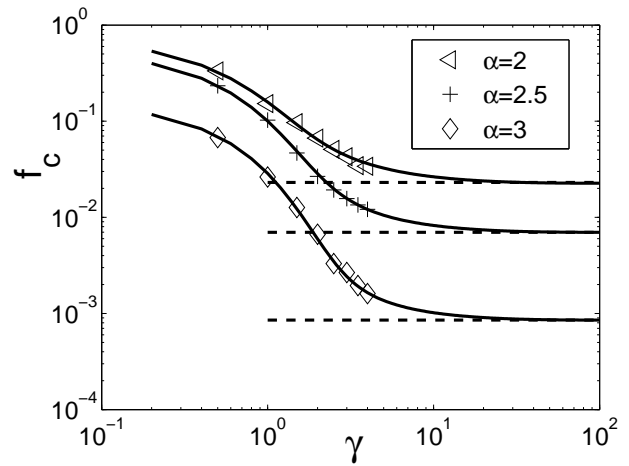}}}
\caption{Percolation
threshold $f_c$ under changes of the attack exponent $\gamma$ for three
different SF networks, $p_k\sim k^{-\alpha}$,  with
$\alpha=2,2.5$ and $3$, and $N=10^5$.
Symbols correspond to stochastic simulations, while solid curves correspond to Eqs.~(\ref{eq:critical_intensity}) and (\ref{eq:percolation_threshold_attack}). The horizontal dashed lines indicate the asymptotic value of $f_c$ given by Eq.(\ref{eq:lim}).
}\label{fig:percolate}
\end{figure}

\vspace{0.1cm} \noindent {\it Generalized attack.}
In the following, we
model the various dynamics (attacks) through a generalized equation
of the form: $f_k = C \, k^{\gamma}$, where $\gamma$ is a real
number signifying the amount of
 network structure information available to the attacker to breakdown the
network~\cite{gallos2005}, and $C$ is a constant that we refer to as attack intensity.
Clearly, $\gamma>0$ represents a situation in which high degree
nodes are removed with higher probability, while $\gamma<0$ models
the opposite. The last case is suitable to situations in which low
degree nodes are more prone to fail.
We are interested in knowing, for a given $\gamma$, the critical
fraction $f_c$ of nodes that is required to remove through such attack in order to destroy the
network, i.e., the percolation threshold.
Thus, the problem reduces to compute, for a given $\gamma$, the critical attack intensity $C^{*}$.
Replacing in Eq.~(\ref{eq:stabilitycond}) the above definition of $f_k$ and after some algebra we obtain:
\begin{eqnarray}\label{eq:critical_intensity}
C^{*}=
 \frac{1}{2 q \langle k^{\gamma+1} \rangle} \left\{ \left[2 \langle k \rangle q  - Q - (1/N) \left( q + \langle k^{\gamma+1} \rangle \right) \right]\right. \\
\nonumber  \left. -\left[ Q^2 + (1/N) \left(2Q + \left(q+ \langle k^{\gamma+1}\rangle\right)^2  \right) - 4qp \langle k^{\gamma+1} \rangle \right]^{1/2} \right\} \, ,
\end{eqnarray}
where $p=\langle k^2 \rangle -2 \langle k \rangle$, $q=\langle k^{\gamma+1} \rangle - \langle k^{\gamma+2} \rangle$, $Q = \langle k^{\gamma+1}\rangle p +\langle k\rangle q$, and $\langle k^{\omega}\rangle$ is defined as $\langle k^{\omega}\rangle = \sum_{k} k^{\omega}\,p_k$.
Since the fraction of removed nodes $f$ after an attack is
$f=\sum_{k} p_k f_k$, the expression for percolation threshold $f_c$ is simply:
\begin{eqnarray} \label{eq:percolation_threshold_attack}
f_c= C^{*}\langle k^{\gamma} \rangle \, .
\end{eqnarray}
Fig.~\ref{fig:percolate} illustrates the behavior of $f_c$ on three
SF networks ($\alpha=2$, $2.5$ and $3$) upon changes in the attack
exponent $\gamma$.
The symbols correspond to stochastic simulations performed on the networks of size $N=10^5$,
while the black curves refer to Eqs.(\ref{eq:critical_intensity}) and (\ref{eq:percolation_threshold_attack}).
In the numerical experiments we have computed $f_c$ following \cite{gallos2005}:
when the fraction of removed nodes is $f_c$, the probability $F$ of finding the network with $\kappa'>2$ is $1/2$.

It is interesting to observe that for any SF network, the minimum
fraction $\Phi_c$ of nodes that is required to be removed to break down
network is obtained by taking the limit $\gamma\to\infty$ of Eq.
(\ref{eq:percolation_threshold_attack}):
\begin{eqnarray}\label{eq:lim}
\Phi_c=\lim_{\gamma\to\infty}f_c(\gamma,\alpha) = h(\alpha)\frac{1}{k_M (k_M -1)} \,
\end{eqnarray}
where $h(\alpha)$ is $h(\alpha)=\langle k^2\rangle - 2 \langle k\rangle$ and $k_M$ is the maximum degree of the original
network.
Notice that Eq.(\ref{eq:lim}) represents an attack performed having full knowledge on the network topology.
The asymptotic values corresponding to Eq.(\ref{eq:lim}) are shown in Fig.~\ref{fig:percolate} as horizontal dashed lines.
Fig.~\ref{fig:percolate} indicates that typically an increase of the
information about the network topology, resp. $\gamma$, helps the
attacker to break down the network with the removal of a smaller
number of nodes. However,  information becomes redundant as the
asymptotic value $\Phi_c$ is approached.

\vspace{0.1cm} \noindent {\it Finite network size.}
To illustrate the effect of network size $N$ upon the percolation
threshold $f_c(N)$, we customize Eq.(\ref{eq:critical_intensity})
for random attack or failure. When $\gamma=0$ and $f_k$ is
independent of $k$, we obtain:
\begin{eqnarray}
\label{eq:percolation_N} f_c(N) = f_c^{\infty} +   \frac{1}{N}
\left(\frac{2-(\langle k^2 \rangle/\langle k \rangle)}{\langle k^2
\rangle-\langle k \rangle}\right) \, ,
\end{eqnarray}
where $f_c^{\infty}$ is the well-known percolation threshold for infinite networks under failure~\cite{cohen00,callaway2000} which reads:
$f_c^{\infty} =  1- \left[1/\left(\langle k^2 \rangle/\langle k \rangle - 1 \right) \right]$.

\begin{figure}
\centering\resizebox{\columnwidth}{!}{\rotatebox{0}{\includegraphics{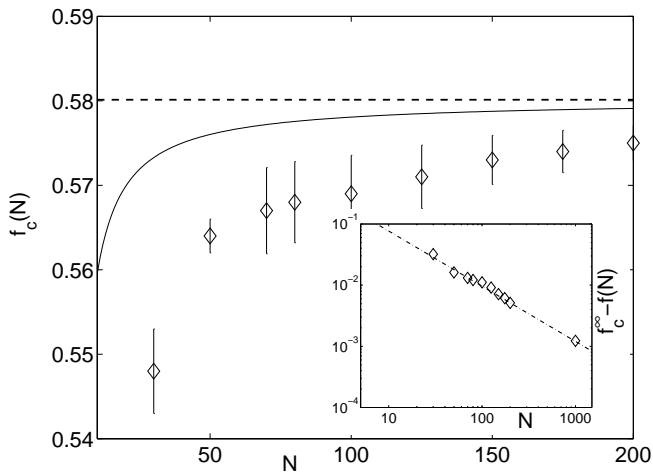}}}
\caption{Percolation threshold $f_c$ as function of $N$.
Symbols correspond to stochastic simulation, solid curve to Eq. (\ref{eq:percolation_N}), and dashed curve to the classical result $f_c^{\infty}$.
The error bars represent the confidence interval that contains $f_c^{\infty}(N)$.
Simulations were performed with ER networks with average degree $\langle k \rangle=3$ and maximum
degree $k_M=5$.
The inset shows the scaling of $f_c^{\infty}-f_c(N)$ with respect to $N$. The best fitting of the data corresponds to a slope $-0.92 \pm 0.05$ (dashed line).
} \label{fig:sim_finite}
\end{figure}

Fig~\ref{fig:sim_finite} shows a comparison between Eq.~(\ref{eq:percolation_N}) (solid line),  $f_c^{\infty}$ (dashed line), and stochastic simulations (symbols) for ER networks of different sizes.
Notice that Eq. (\ref{eq:percolation_N}) predicts the correct scaling of $f_c$ with $N$, i.e., $f_c(N)-F^{\infty}_{c} \sim N^{-1}$.
The observed deviation between Eq.~(\ref{eq:percolation_N}) and simulations can be arguably attributed to correlations effects, which have been ignored in the current approach.

\vspace{0.1cm} \noindent {\it Concluding remarks.} We
have proposed a general procedure to calculate the distorted degree
distribution of uncorrelated finite size networks under arbitrary
failure/attack.
Using the expression for the distorted degree distribution we have derived  the critical
condition for the stability of  finite size networks.
The formalism has been further applied to derive an expression for the percolation threshold under a general attack.
Finally, it was shown that  the obtained percolation threshold predicts  an additive correction
which scales as $N^{-1}$ with respect to the classical result for infinite networks, as observed in simulations.

The results derived throughout this manuscript are valid only for uncorrelated networks.
The effect of correlations on the percolation threshold for finite (and infinite) networks
remains as one of the major challenges. Further extensions of this theory will be focused in that direction.

The authors thank H.Chat\'e, F.Ginelli, L.Brusch and A.Deutsch for insightful comments on the
manuscript. FP acknowledges the hospitality of IIT-Kharagpur.

\end{document}